\documentclass[twoside]{ilcws08}
\usepackage[latin1]{inputenc}
\usepackage[dvips]{graphicx,epsfig,color}
\usepackage{wrapfig,rotating}
\usepackage{amssymb,amsmath,array}

\begin{document}
\title{
%%%%   Paper title goes here  %%%%%%%%%%%%%%
Development of MicroMegas for a Digital Hadronic Calorimeter}
%***********************************************************************
% AUTHORS INFORMATION AREA
%***********************************************************************
\author{Catherine Adloff, Jan Blaha$^{\dag}$, Ambroise Espargilière and Yannis 
  Karyotakis
% Optional short acknowledgment: remove next line if non-needed
\thanks{This work was performed within the CALICE collaboration.}
% DO NOT MODIFY THE FOLLOWING '\vspace' ARGUMENT
\vspace{.3cm}\\
% Addresses and institutions (remove "1- " in case of a single institution)
University of Savoie and the Laboratoire d'Annecy-le-Vieux de Physique 
des Particules \\
Annecy-le-Vieux, F-74940 France\\
\\
\dag \ - Contact: \tt{\bf jan.blaha@lapp.in2p3.fr}\\
%% Remove the next three lines in case of a single institution
%\vspace{.1cm}\\
%2- Second Author's Institution - Department \\
%Address of Second Author's Institution - Country\\
}
%%***********************************************************************
% END OF AUTHORS INFORMATION AREA
%***********************************************************************

\maketitle

\begin{abstract}
Recent developments on the MicroMegas prototypes built by use of the bulk 
technology with analog and digital readout electronics are presented.
The main test beam results of a stack of several MicroMegas prototypes
fully comply with the needs of a hadronic calorimeter for future particle 
physics experiments. A technical solution for a large scale prototype 
is also introduced. 
\end{abstract}

\section{Introduction}
Future particle physics experiments at the International Linear Collider 
(ILC)~\cite{ilc} will employ the Particle Flow Algorithm (PFA). In order to 
achieve an optimal PFA performance, a highly granular hadronic calorimeter with 
a good shower separation is required. One of the suitable and affordable choice
for an active part of the hadronic calorimeter is a thin gaseous detector with 
embedded {\it digital} (1-bit) or {\it semi-digital} (2-bit) readout. This 
concept allows the construction of the so-called Digital Hardronic CALorimeter 
(DHCAL) with very fine granularity (a cell size of about 1~cm$^2$) providing 
high MIP efficiency, low hit multiplicity as well as negligible performance 
degradation due to high dose rates, hadronic showers and aging. 

One of the promising candidate for a DHCAL is the MICRO MEsh GAseous Structure
(MicroMegas) which is a micro-pattern gaseous detector~\cite{micromegas}. 
Prototypes developed at LAPP consist of a commercially available 20~$\mu$m thin
woven mesh which separates the 3~mm drift gap from the 128~$\mu$m amplification
gap. This simple structure allows full efficiency for MIPs and provides a good 
gain uniformity over the whole detection area. Due to the fast collection of 
the amplification charge, the MicroMegas counting rate is very high and not 
constrained as in the case of the Glass RPC. Moreover, the tiny size of the 
amplification avalanche results in fast signals without physical cross talk 
and, consequently, low multiplicity. The chosen bulk technology based on 
industrial PCB processes, offers a robust large area detector with working 
voltages lower than 500~V. The MicroMegas with 1~cm$^2$ anode pads is
therefore a very appealing possibility to equip a DHCAL well optimized for 
the PFA. 

\section{MicroMegas Prototypes}
Three different kinds of MicroMegas prototypes with 1~cm$^2$ pads, were 
developed and built at LAPP. The first type is equipped with analog readout
and the two others with embedded digital readout ASICs. 
   
The analog readout, intended for full detector characterization, uses 16-channel
GASSIPLEX chips connected to a 12~bit VME ADC, which provides charge 
determination with a high resolution (0.4 fC/ADC~Count). The data acquisition
is performed by the CENTAURE program~\cite{centaure}. Three MicroMegas with 
$6\times12$ pads and one with $12\times32$ pads were equipped with this analog 
readout.

\begin{figure}[htb]%{0.5\columnwidth}
\centerline{\includegraphics[width=0.5\columnwidth]{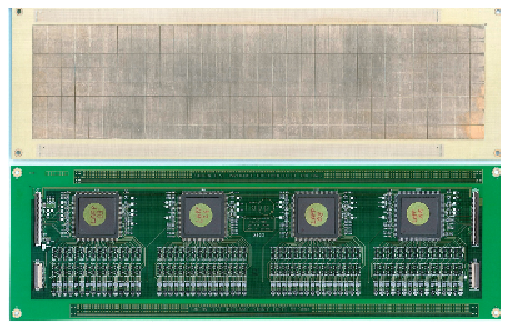}\hfill
  \includegraphics[width=0.32\columnwidth]{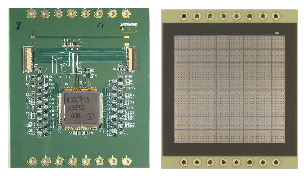}}
\caption{MicroMegas prototype with HARDROC (left) and DIRAC (right) readout.
  Both photographs show the pad and ASICs sides.}\label{Fig:prototypes}
\end{figure}

Two mixed-signal ASICs are designed for the digital readout, the 
HARDROC~\cite{hardroc} and DIRAC~\cite{dirac}. The former was chosen as a 
baseline for the 1~m$^3$ European DHCAL project in order to ensure the 
availability of the digital readout of either MicroMegas or Glass RPC at short 
time. Whereas, the latter is a long-term R\&D which aims to obtain a low cost 
ASIC with an easy signal routing implementation on the detector PCB, simple 
calibration and digital readout down to MicroMegas MIP charges. Four MicroMegas
with $8\times32$ pads (see Fig.~\ref{Fig:prototypes} left) and one with 
$8\times8$ pads (see Fig.~\ref{Fig:prototypes} right) were built with HARDROC
and DIRAC readout, respectively. 

All MicroMegas bulks are realized by lamination at high temperature of 
photosensitive foils and a mesh laid on a PCB with different signal routing 
depending on the readout. By photo-lithography, the photosensitive foils are 
etched producing the 128~$\mu$m pillars. A thin copper foil, glued to part of 
the calorimeter absorber medium (2~mm thick plate out of a 2 cm thick 
absorber), defines the drift cathode. The top of the chamber is therefore not 
contributing to the active medium thickness. The drift gap is realized with a 
3~mm thick frame which provides also the gas inlets and outlets.

\section{X-ray response}
In order to determine the gain and energy resolution, each prototype was
exposed to an $^{55}$Fe X-ray source. The gain with the analog readout was 
measured up to 10000 and the energy resolution down to 8.5\% corresponding 
to a FWHM of 19.6\%. The gain and FWHM were also measured as a function of 
the drift and amplification fields, gas flow and pressure. The expected
exponential dependence on the gain on the amplification field was verified 
and an absolute pressure dependence of -2~fC/mbar was determined.

%\begin{figure}[htb]%{0.5\columnwidth}
%\centerline{\includegraphics[width=0.48\columnwidth]{xray.eps}\hfill
%  \includegraphics[width=0.48\columnwidth]{xray2.eps}}
%\caption{X-ray response.}\label{Fig:Xray}
%\end{figure}

\section{Test beam results}
The behavior of the MicroMegas prototypes under high energy particle 
irradiation was studied during two test beam periods at CERN. In the first 
period, four prototypes with analog readout and one prototype with DIRAC 
digital readout were assembled in a stack and tested in 200~GeV muon and pion 
beams at the H2 SPS line. In the second period, four prototypes with digital 
readout based on the HARDROC chip with associated
electronics were exposed to 7~GeV pions at the T9 PS line. Since the analysis
of data collected during the second period is ongoing, only results obtained 
during the first period in summer 2008 are presented below.

\begin{figure}[htb]%{0.5\columnwidth}
\centerline{\includegraphics[width=0.48\columnwidth]{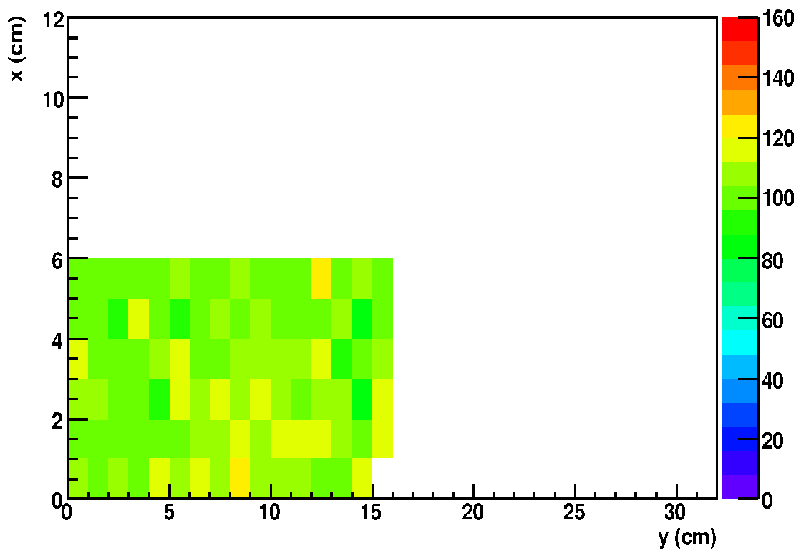}\hfill
  \includegraphics[width=0.48\columnwidth]{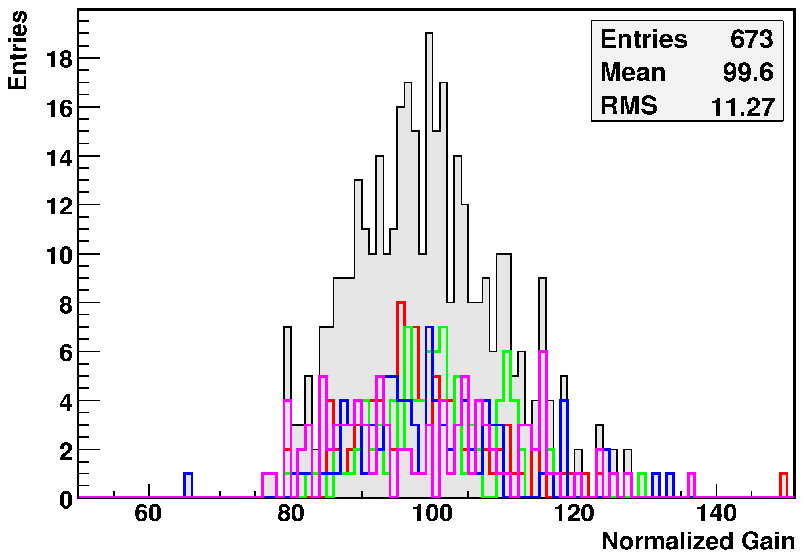}}
\caption{Map of the MPV of deposited energy distribution for one chamber (left)
  and the relative MPV distributions (right) for four MicroMegas prototypes 
  with analog readout.}
\label{Fig:MIP}
\end{figure}

The mapping of each chamber was performed in terms of pedestal and electronic 
noise, Most Probable Value (MPV) of deposited energy distribution and its 
standard deviation. The pedestal Gaussian distribution showed very good noise 
performance with an average noise of 0.6 fC. A clear Landau distribution
of deposited energy was obtained for each pad with a MPV around 45~fC. The MPV 
is well uniform for all the chambers (see Fig.~\ref{Fig:MIP} left) with an 
average dispersion of 11\% RMS (see Fig.~\ref{Fig:MIP} right).

For efficiency and multiplicity studies, the hit threshold was set to 
27~ADC~Counts (2.8~fC). The counted 
hit in a chamber is considered only in case when three hits of three other 
chambers are sitting on the extrapolated straight line with respect to the 
chosen chamber. The efficiency of each pad is then computed as the ratio of 
the counted number of hits in the central or adjacent pad, so in a $3\times3$ 
pad array, and the expected number of hits. The efficiency varies from 
$(92,99 \pm 0,10)\%$ to $(97,05 \pm 0,07)\%$ for the oldest and newest 
prototype showing considerable progress in the manufacturing process. The hit 
multiplicity is measured by counting the number of hits in the same pad array. 
A multiplicity smaller than 1.1 was found for all of the four prototypes. 

In the same test beam period, the first prototype with digital readout using 
DIRAC ASIC was tested in a 200 GeV pion beam. The functionality of the 
prototype was verified by beam scanning across the chamber. Nevertheless, 
further tests with a stack of several prototypes are compulsory to measure 
threshold dependence, efficiency and multiplicity. These tests are foreseen 
for spring 2009.

%\begin{wraptable}{l}{0.5\columnwidth}
%\centerline{
%\begin{tabular}[htb]{|l|c|}
%\hline
%Prototype 0 (96 pads)  & $(97,05 \pm 0,07)\%$ \\
%Prototype 1 (96 pads)  & $(98,54 \pm 0,05)\%$ \\
%Prototype 2 (96 pads)  & $(92,99 \pm 0,10)\%$ \\
%Prototype 3 (384 pads) & $(96,17 \pm 0,07)\%$ \\
%\hline
%\end{tabular}
%}
%\caption{Efficiency}
%\label{tab:Efficiency}
%\end{wraptable}

\section{Future developments}
\begin{figure}[htb]%{0.5\columnwidth}
\centerline{\includegraphics[width=0.51\columnwidth]{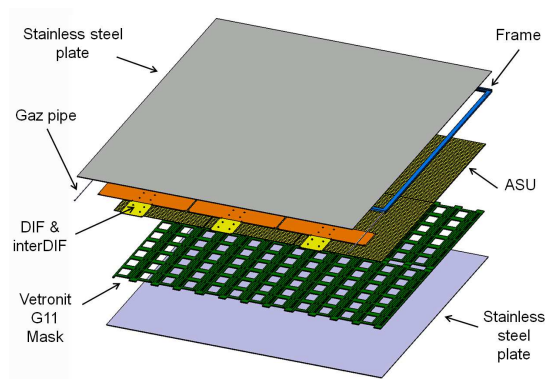}}
\caption{1m$^2$ MicroMegas prototype}
\label{Fig:1m2}
\end{figure}

The next step is the development of a 1~m$^2$ MicroMegas prototype with 9216 
readout pads. The 1~m$^2$ is an assembly of six Active Sensor Units
(ASUs) with 24 ASICs each closed by two plates of 2 mm thick stainless 
steel (see Fig.~\ref{Fig:1m2}). In order to avoid destructive sparks from the 
increased capacitance of a too large area mesh, one mesh per ASU will be used. 
The total thickness of the prototype should not exceed 6~mm (without absorber).
The construction of such a 1~m$^2$ prototype is scheduled for the beginning 
of 2009 and its first test in a beam for late 2009. The 1~m$^2$ design is 
foreseen for large quantity production in order to build a 1 m$^3$ DHCAL 
prototype.

%\begin{wrapfigure}{r}{0.5\columnwidth}
%\end{wrapfigure}

\section{Summary and Conclusion}
Several MicroMegas prototypes with analog and digital readout have been 
successfully built and tested. The summer 2008 test beam results 
have shown very good performance complying with the DHCAL requirements. The
first operational bulk MicroMegas with embedded digital electronics was 
realized and exposed to a pion beam. Further studies with a stack of these new
thin prototypes are foreseen. Development of a large scale prototype 
compatible with the CALICE DAQ is well underway to be ready for a test beam 
in~2009.

%\section{Acknowledgments}
%We would like to thank all the colleagues who directly or indirectly have 
%contributed to the MicroMegas developments and also those who have been 
%participating in the test beams at CERN.

%\section{Bibliography}

%\subsection{Link to slides}

% ****************************************************************************
% BIBLIOGRAPHY AREA
% ****************************************************************************

\begin{footnotesize}
% IF YOU DO NOT USE BIBTEX, USE THE FOLLOWING SAMPLE SCHEME FOR THE REFERENCES
% ----------------------------------------------------------------------------

% ----------------------------------------------------------------------------

% IF YOU USE BIBTEX,
% - DELETE THE TEXT BETWEEN THE TWO ABOVE DASHED LINES
% - UNCOMMENT THE NEXT TWO LINES AND REPLACE 'Name_Of_Your_BibFile'

%\bibliographystyle{unsrt}
%\bibliography{Name_Of_Your_BibFile}
% example of Name_Of_Your_BibFile.bib
% @Article{Turcato:2006ch,
%      author    = "Turcato, M.",
%  collaboration = "ZEUS and H1",
%      title     = "Lepton flavour violation and charmonium physics at HERA",
%      journal   = "Nucl. Phys. Proc. Suppl.",
%      volume    = "162",
%      year      = "2006", 
%      pages     = "283-287",
%      SLACcitation  = "%%CITATION = NUPHZ,162,283;%%"
% }
% 
% @Unpublished{Gogitidze:2007du,
%      author    = "Gogitidze, N.",
%  collaboration = "H1", 
%      title     = "Prompt photons and particle momentum distributions at
%                   HERA", 
%      year      = "2007",
%      note    = "hep-ex/0701033",
%      SLACcitation  = "%%CITATION = HEP-EX 0701033;%%"
% }

\end{footnotesize}

% ****************************************************************************
% END OF BIBLIOGRAPHY AREA
% ****************************************************************************

\end{document}